%% file: hfqc.tex
\documentclass{aa}    
\pdfoutput=1
\usepackage{graphicx,twoopt}
\usepackage[varg]{txfonts}           

\usepackage{pdfcomment}              
\usepackage{acronym}                 

\usepackage{url}

\usepackage{color,hyperref}

\definecolor{linkcolor}{rgb}{0,0,0.5}
\hypersetup{colorlinks=true,
					linkcolor=linkcolor, 
					citecolor=linkcolor,
                    filecolor=linkcolor, 
                    urlcolor=linkcolor}

\usepackage{natbib}
\bibpunct{(}{)}{;}{a}{}{,} 

\newcommand{\figref}[1]{\ref{fig:#1}}
\newcommand{\Fig}[1]{\figurename~\figref{#1}}
\newcommand{\fig}[1]{\Fig{#1}}
\newcommand{\figlabel}[1]{\label{fig:#1}}
\newcommand{\Tab}[1]{Table~\ref{tab:#1}}
\newcommand{\tab}[1]{\Tab{#1}}
\newcommand{\tablabel}[1]{\label{tab:#1}}
\newcommand{\Eq}[1]{Equation~(\ref{eq:#1})}
\newcommand{\eq}[1]{\Eq{#1}}

\newcommand{\eqlabel}[1]{\label{eq:#1}}
\newcommand{\sectionname}{Sect.}
\newcommand{\Sect}[1]{\sectionname~\ref{sect:#1}}
\newcommand{\sect}[1]{\Sect{#1}}
\newcommand{\sectionnamei}{Section}
\newcommand{\Secti}[1]{\sectionnamei~\ref{sect:#1}}
\newcommand{\secti}[1]{\Secti{#1}}

\newcommand{\App}[1]{Appendix~\ref{sect:#1}}
\newcommand{\app}[1]{\App{#1}}
\newcommand{\sectlabel}[1]{\label{sect:#1}}

\newcommand{\lya}{Ly$\alpha$} 
\newcommand{\lyb}{Ly$\beta$} 
\newcommand{\hb}{H$\beta$} 
\newcommand{\ha}{H$\alpha$}

\newcommand{\oii}{[\ion{O}{ii}]$\lambda$3727} 
\newcommand{\oiii}{[\ion{O}{iii}]$\lambda$5007}
 
\newcommand{\feii}{\ion{Fe}{ii}} 
\newcommand{\civ}{\ion{C}{iv}} 
\newcommand{\mgii}{\ion{Mg}{ii}} 

\newcommand{\fluxunit}{erg s$^{-1}$ cm$^{-2}$ \AA$^{-1}$}

\begin{document}  
\title{An X-shooter composite of bright $1<z<2$ quasars from UV to
  infrared \thanks{Based on observations made with telescopes at the
    European Southern Observatory at La Silla/Paranal, Chile under
    program 090.A-0147(A).}, \thanks{The quasar composite is available in electronic form
    at the CDS via anonymous ftp to \href{http://cdsarc.u-strasbg.fr/}{cdsarc.u-strasbg.fr} (\href{http://130.79.128.5}{130.79.128.5}) 
    or via \url{http://cdsweb.u-strasbg.fr/cgi-bin/qcat?J/A+A/}. Source code and composite is also made public at \url{https://github.com/jselsing/QuasarComposite}.}}

\titlerunning{X-shooter quasar composite}

\author{J. Selsing\inst{1}, J.~P.~U. Fynbo\inst{1}, L. Christensen\inst{1},
J.-K. Krogager\inst{1}
		  }

\authorrunning{J. Selsing et al.}

\institute{
	Dark Cosmology Centre, Niels Bohr Institute, University of Copenhagen, Juliane
Maries Vej 30, 2100 Copenhagen, Denmark. 
		 }

\abstract{Quasi-stellar object (QSO) spectral templates are important both to QSO
  physics and for investigations that use QSOs as probes of
  intervening gas and dust. However, combinations of various QSO
  samples obtained at different times and with different instruments so as
  to expand a composite and to cover a wider rest frame wavelength region may
  create systematic effects, and the contribution from QSO hosts may
  contaminate the composite. We have constructed a composite spectrum 
  from luminous blue QSOs at $1< z < 2.1$ selected from
  the Sloan Digital Sky Survey (SDSS).  The observations with X-shooter simultaneously cover ultraviolet (UV)
  to near-infrared (NIR) light, which ensures that the composite
  spectrum covers the full rest-frame range from Ly$\beta$ to 11350
  {\AA} without any significant host contamination. Assuming a power-law continuum for the composite we find a spectral slope of $\alpha_\lambda$
  = 1.70$\pm$0.01, which is steeper than previously found in the
  literature. We attribute the differences to our broader spectral wavelength coverage, which allows us to effectively
  avoid fitting any regions that are affected either by strong QSO emissions lines
  (e.g., Balmer lines and complex [Fe II] blends) or by intrinsic host
  galaxy emission. Finally, we demonstrate the application of the
  QSO composite spectrum for evaluating the reddening in other
  QSOs. }


\maketitle
%

\section{Introduction}     \sectlabel{introduction}
Template spectra built as composites of many carefully selected
individual spectra are useful for a wide range of purposes, e.g., detecting features that are too weak to be detected in individual
spectra, identifying objects that differ from the template, using
in modeling/spectral energy distribution (SED) fitting, etc. Examples of such composite spectra
include template spectra of various classes of galaxies
\citep{Mannucci2001, Shapley2003, Dobos2012}, QSOs
\citep{CristianiS.andVio1990, Boyle1990, Francis1991, Zheng1997,
  Brotherton2000, VandenBerk2001, Telfer2002, Richards2006a,
  Glikman2006, Lusso2015} and gamma-ray burst (GRB) afterglows \citep{Christensen2011}.
When investigating dust in QSOs or in galaxies along QSO sightlines, template spectra can be used to determine the amount of
extinction by artificially reddening the template to match an observed
spectrum
\citep[e.g.,][]{Glikman2007,Urrutia2009,Wang2012,Fynbo2013,Krogager2015}.
The amount of extinction inferred will depend on the assumed
extinction curve. Many studies find that Small Magellanic Cloud (SMC)-like extinction provides
a good match \citep{Richards2003,Hopkins2004}. In some cases steeper
extinction curves are required
\citep{Fynbo2013,Jiang2013,Leighly2014}.

The wavelength coverage of a single instrument is always limited, and
therefore templates from different telescopes sample different parts
of the SED. The redshift distribution of QSOs
are exploited to extend the wavelength coverage of a single instrument,
rejecting wavelength regions affected by strong atmospheric
absorption, but as a consequence, different wavelength regions of the
template will be constructed from differing redshift
intervals. Several instruments can be combined to extend the
wavelength coverage of a single template as is done in
\citet{Glikman2006} where SDSS \citep{Gunn2006} and IRTF, SpeX
\citep{Rayner2003} are combined to cover the range from 2700 \AA~ to
37000 \AA, or several templates can be combined to cover regions of
interest as done in \citet{Zhou2010}, where the template of
\citet{VandenBerk2001} is stitched together at 3000 \AA~with the
composite by \citet{Glikman2006}. 

The template by \citet{VandenBerk2001} is widely used in studies
involving QSOs, where it has been very useful as a cross-correlation
template to determine redshifts \citep{Stoughton2002, Rafiee2011}, as
a model for the optical QSO SEDs \citep{Croom2004, Hopkins2006,
  Hopkins2007}, and for studies of the host galaxies of AGN
\citep{Kauffmann2003b}. When using the \citet{VandenBerk2001} template,
it is important to know that it contains significant host galaxy
contamination, in particular at redder wavelengths \citep[e.g.,][their
  Fig.~5]{Fynbo2013}.  In this paper, we use data from the X-shooter
spectrograph to build a new QSO template that covers the full range
from rest-frame Ly$\beta$ to 11350 {\AA} based on bright SDSS QSOs and
observed with a single instrument over this spectral range. 

In \sect{sample} we describe the sample selection and provide details
of the observations. \secti{construct} describes the construction of
the composite spectrum, and \sect{results} presents the resulting
composite. In \sect{discuss} we perform a comparison with existing
composites and in \sect{conclusion} offer our conclusion. We use the
cosmological parameters from \citet{Planck2014} with $H_{0} = 67.77$
km s$^{-1}$ Mpc$^{-1}$, $\Omega_{M} = 0.3071$, and $\Omega_{\Lambda} =
0.6914$ throughout the entire paper.

\section{Issues with the currently most used template}   \sectlabel{problem}

The currently most widely used QSO template is the SDSS template of
\citet{VandenBerk2001}. As described very clearly in that paper, the
SDSS composite spectrum shows a significant change to a shallower
spectral slope around 5000 \AA. This is mainly attributed to
contaminating light from the underlying host galaxies. Other effects
also contribute, such as the emission from a hot dust component, but the
dominating factor is probably the host contamination. Our main
motivation for building a QSO template without this problem is the
following: A search for dust-reddened QSOs using
near-IR (NIR) selection of QSOs has been initiated \citep{Fynbo2013, Krogager2015}. 
In the central southern SDSS footprint, Stripe 82 \citep{Annis2014}, about 50 such candidate QSOs have been studied using the
New Technology Telescope (NTT). 

Although the search was for QSOs reddened by foreground absorber
galaxies, most systems turned out to be QSOs reddened by dust in their
host galaxies. The optical spectra can be matched well by the SDSS
template spectrum reddened by SMC-like extinction curve, but the NIR
(rest frame optical) photometry from the UKIRT Infrared Deep Sky Survey (UKIDSS) cannot be simultaneously
fitted. The problem is illustrated in Fig. 6 in \citet{Fynbo2013},
where they attempt to model the EFOSC2 spectrum of a dust-reddened QSO
at z = 1.16 observed in that survey. As seen, the SDSS spectrum and
photometry can be matched well with the Vanden Berk template, but the
photometry from UKIDSS is much too blue for even the unreddened
template. This is not a unique case, but a problem that is seen for
all the dust-reddened QSOs found in their search.  In the SDSS
composite, the QSOs which contribute to the spectrum at $ > 5000$
\AA~have to be at fairly low redshifts ($z \lesssim 0.5$). These QSOs
have absolute magnitudes that are three to four magnitudes fainter than bright
$z > 1.5$ QSOs \citep[e.g.,][their Fig.~1]{VandenBerk2001}. They are
also likely to have the brightest rest frame optical host galaxies, so host contamination is not surprising. 

By selecting intrinsically
brighter QSOs at higher redshifts we should be able to avoid the
problem with host contamination.  To circumvent the problem
of host galaxy contamination, \citet{Fynbo2013} and
\citet{Krogager2015} stitched together template spectra from different
authors, which increases the wavelength coverage and avoids host galaxy
contamination, but by extension introduces different selection
effects. The composite by \citet{Glikman2006} is stitched together
with the one by \citet{VandenBerk2001} at $3000$ \AA~and covers the
entire range from $0.08 - 3.5 \mu m$ with a single template, but is
constructed with different instruments and an intrinsic differing QSO
sample across the entire wavelength range.  What we present here is a
small, well-defined sample selected to cover from \lyb~to \ha~and the
entirety of the "Small Blue Bump" to effectively determine the
continuum level on both sides, which is crucial for normalizing
the input spectra. This will effectively also allow us to determine
the slope of the far-UV better by having access to more uncontaminated
continuum, simply because of the increased wavelength coverage.

\section{Sample description}   \sectlabel{sample}

In order to build a QSO spectrum with no significant host galaxy
contamination we select very bright ($r \lesssim 17$) SDSS QSOs at
redshifts $1 < z < 2.1$ for which host galaxy contamination should be
insignificant and where we can use the redshift distribution to cover
the regions of strong telluric absorption. Previous QSO templates in
the literature have been based on hundreds of spectra, but this is
neither possible nor necessary for us. We are mainly interested in
tracing the shape of the continuum with negligible host contamination
and wide wavelength coverage and for that a smaller sample will
suffice. From the 166,583 quasars presented in the SDSS DR10 quasar catalog \citep{Paris2014}, 173 objects fulfill our selection criteria. Of these quasars, 102 have declinations $\lesssim +15^\circ$ from which we select seven without signs of BAL activity or other strong associated absorption systems. The selected quasars have SDSS spectra and are observable in a single visitor run at the Very Large Telescope (VLT). With this sample, all rest frame
wavelengths from \lya~to 8500 \AA~will be covered by at least four
spectra. For all of the spectra we cover the region between \oii~and
\oiii. This is useful for determination of precise systemic
redshifts. The targeted QSOs, their coordinates and redshifts are
listed in \tab{targs}.

Spectroscopic observation of the targets have been carried out using X-shooter
\citep{Vernet2011}, the single-object cross-dispersion echelle spectrograph at
the VLT. The observations were carried out 14 -- 16 March, 2013 under ESO
programme 090.A-0147(A). Each observation consists of 4 exposures in nodding mode in
the sequence ABBA with a total integration time of 1800s per object,
simultaneously in the UVB(3100 - 5500~\AA), VIS(5500 - 10150~\AA)  and NIR(10150
- 24800~\AA) spectroscopic arms. The nominal resolving power $R = \lambda /
\Delta \lambda$ for our observations is 4350 in the UVB-arm for a slit width of
1.0 arcsec, 7450 and 5300 in the VIS- and NIR arms respectively, for a slit
width of 0.9 arcsec. For one observation, slit widths of 1.3 arcsec in UVB and
1.2 arcsec for VIS and NIR have been used thus lowering the nominal resolution.
The conditions were photometric, reaching a seeing of 0.66 arcsec as determined
from the width of the trace at 7825 \AA~and therefore the delivered effective
resolution will be higher than tabulated. We determine the instrumental spectral
FWHM using MOLECFIT \citep{Smette2015, Kausch2015} where a model atmosphere is
convolved with a Gaussian kernel and a fit of the telluric absorption bands in
the visual arm is done, where residuals are minimized with respect to the
width of the Gaussian kernel. The size of the kernel used increases linearly
with wavelength to keep the resolution constant, as is the case for X-shooter.
The FWHM values are given relative to the central wavelength of the arm, which
in the case for the visual arm is at 7825~\AA, which we convert to resolving
power.
Our observations are executed under excellent seeing conditions with
corresponding resolutions between 11000 and 14500 in the visual arm with an
average R$_{VIS}$ = 12750 - a factor of 1.7 increase compared the nominal
resolution. 

The data were reduced with the ESO/X-shooter pipeline v2.5.2
\citep{Modigliani2010} using the Reflex interface \citep{Freudling2013}. The
spectra have been rectified on a grid with 0.2 \AA/pixel in the UVB and VIS arm
and 0.6 \AA/pixel in the NIR arm, thus slightly oversampling even the
highest-resolution spectra while minimizing the correlation between adjacent
pixels in the rectification. The spectra were extracted by simple integration in
the nodding window of the rectified image and were flux calibrated using
spectrophotometric standards \citep{Vernet2010, Hamuy1994}. The seeing during the
flux-standard star observations was significantly below the 5 arcsec slit width
used for the observations of the spectrophotometric standard stars and slit
losses will therefore be negligible. To check the accuracy of the flux
calibration the corresponding spectra observed by SDSS \citep{Ahn2014} are
compared to the X-shooter observations. Slight variations in individual spectra
are detected in both flux level and apparent slope at the $\sim$10 per cent
level. This can both be attributed to erroneous flux calibration in one of the
spectra or intrinsic object variability, which is observed to be of this order
\citep{MacLeod2012, Morganson2014}. We discuss this in more detail in
\sect{systematics}.

\input{tabs/targets.tex}

\subsection{Telluric correction}   \sectlabel{telluric}

All ground-based observations suffer from atmospheric absorption, which
is especially true in the Visual (VIS) and Near-Infra-Red (NIR) arms
of X-shooter where there are regions of telluric absorption from water
vapor and other molecules with almost zero transmission. Because of
the limited sample-size, it is desirable to avoid simply masking out
regions of atmospheric absorption, but rather correct for it. To this
end a method based on the method employed in \citet{Chen2014} is
developed.  To correct for the absorption an exact measure of the
transmission of the atmosphere at the time of the observation is
needed. The amount of absorption depends heavily on the exact
conditions through the atmosphere and these change on very short
timescales. Observations of hot B-type dwarf stars are taken within 3
hours of the QSO observations, which are then used to
generate the atmospheric transmission spectrum.

\begin{figure}[t!]
  \centering
  \includegraphics[width=0.99\columnwidth]{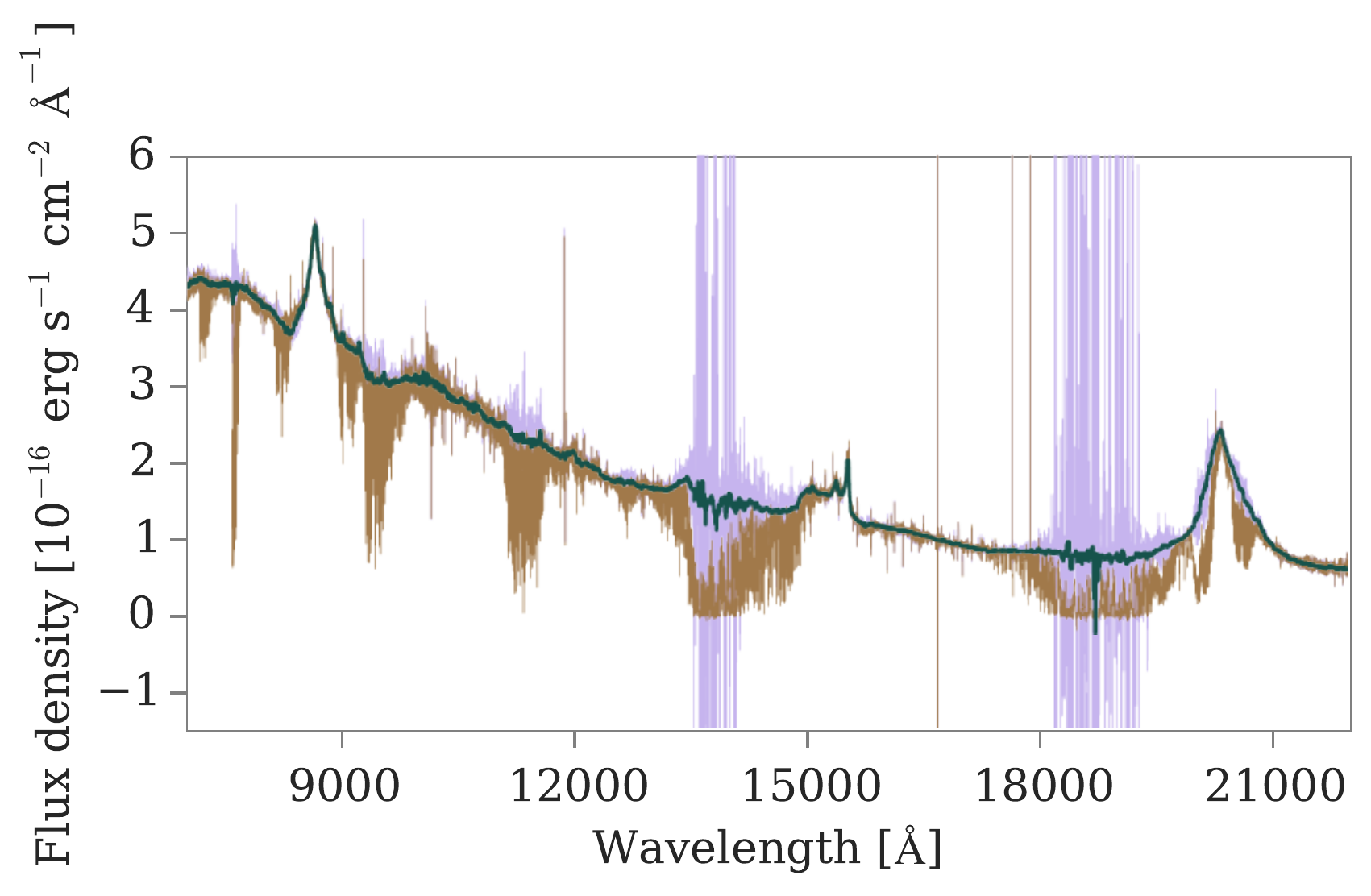}
  \caption[]{Telluric correction for quasar SDSS1431+0535 where brown is the
uncorrected spectrum and purple is the corrected spectrum and teal is the corrected
spectrum smoothed by 50 pixels. The regions of pure noise are where the atmospheric
transmission is $\sim$ zero, thus only leaving noise after correction. Residuals
after correction are visible. The noise image is corrected correspondingly and
regions severely affected are down-weighted in the weighted combination.}
\figlabel{telluric_qc}
\end{figure}

From observations of a telluric standard star it is possible to calculate
the atmospheric transmission spectrum by modeling the intrinsic stellar
spectrum. The transmission spectrum is the fractional difference between the
model and the observed spectrum. We model the telluric standard by finding the
optimal linear combination of templates among the model atmospheres from the
G\"ottingen Spectral Library \citep{Husser2013}. Here the PHOENIX model stellar
atmosphere code is used to create a grid of synthetic spectra in terms of
effective temperature, metallicity and alpha element enhancement. To find the
optimal template we simultaneously fit for the instrumental profile broadening,
velocity shift and optimal template using the penalized-pixel fitting software
pPXF \citep{Cappellari2004}. For the fitting we exclude all regions of strong
telluric absorption since we only want to trace the stellar features. After a
transmission spectrum has been constructed from the telluric standard
observations, the object observation closest in time is divided by the
transmission, thus correcting the absorption. As can be seen in
\fig{telluric_qc}, the intrinsic object spectrum is recovered relatively well.
The regions where the continuum of the telluric standard is not found accurately
will produce an erroneous transmission correction and introduce an error in the
resulting corrected spectrum. We review the consequences of this in
\sect{systematics}.

\section{Composite construction}   \sectlabel{construct}

There is no unique way to construct a composite and the final result is
influenced by the method employed. In the following we describe how the
composite is constructed in this work.

\subsection{Determining the redshift}  \sectlabel{redshifts}

Of crucial importance for the accuracy of the composite is
determination of precise redshifts to accurately move the objects to
the rest frame. Getting the systemic redshift of quasars from emission
lines is complicated by the complex physical structure of the broad
line regions in which the broad emission lines are emitted. For the
bright quasars selected for our composite many of the prominent
high-ionization lines visible will arise in hot clouds with high
peculiar velocities and will therefore be affected by systematic
line shifts \citep{Tytler1992, Richards2002b, Gaskell2013}. To get a
redshift closer to the systemic we choose \oiii, arising in the
narrow-line region and therefore less affected by systematic offsets
\citep{Hewett2010}. \oiii~is situated on a broad Fe-complex and
slightly blended with [\ion{O}{iii}]$\lambda$4959 and \hb, with \hb~containing both broad and narrow components.  
This makes it difficult to
model the entire emission complex by the three lines alone and
therefore only the narrowest component of the \oiii-line is used for
the fit. This is similar to what is done in \citet{VandenBerk2001}.
We mark both sides of the line belonging to
\oiii~where it starts to become blended, fit a low order polynomial
to the edges and subtract the estimated pseudo-continuum.
We define a Gaussian in terms of the rest wavelength of \oiii~and
leave the redshift as a free parameter and fit to the selected
region where a weighted minimization of the residuals is done using
least squares. The best-fit value and the confidence intervals on the
best-fit parameters is determined by resampling the spectrum within
the errors and repeating the continuum estimation and the fit on the
resampled spectrum 10000 times. The weights used are the inverse
variance, $ w = \frac{1}{\sigma^2}$, where $\sigma$ is the associated
error spectrum. The 1$\sigma$ confidence intervals on the fit
parameters are set by the 16th and 84th percentile of the resampling
realizations. Least squares minimization does not guarantee that the global minimum of the
 residuals will be found, but we check
visually that the fits are satisfactory. \oiii~ is visible in all of
our spectra in varying strength so additional refinement is not
needed. As a starting guess for the redshift we use redshifts queried
from SDSS and all fits are visually inspected to check if the fitted
redshift is an improvement on the position of \oiii. We show a fit to
\oiii~ in \Fig{linefit}, where \hb, [\ion{O}{iii}]$\lambda$4959 and
\oiii~ are clearly detected and marked on the plot with vertical
dashed lines. We transform all redshifts to barycentric standard of rest for consistent
comparison with SDSS. For all of our spectra we find slight
corrections to the SDSS redshifts, based on the fit to \oiii, as is
also shown in \tab{targs}. We note that only one out of seven
redshifts agree within the errors, but stress that the measured
redshifts are not measured in a consistent manner and therefore
offsets are expected. SDSS spectral redshift determinations are based
on rest frame UV lines, which are known to exhibit substantial velocity
shifts relative to the systemic redshift \citep{Tytler1992,
  Hewett2010}. The wavelength calibration is accurate to within 1
pixel \citep{Kruhler2015}, which translates into a redshift error
$\lesssim 3\times 10^{-5}$, less than reported errors.

\begin{figure}[t!]
  \centering
  \includegraphics[width=0.99\columnwidth]{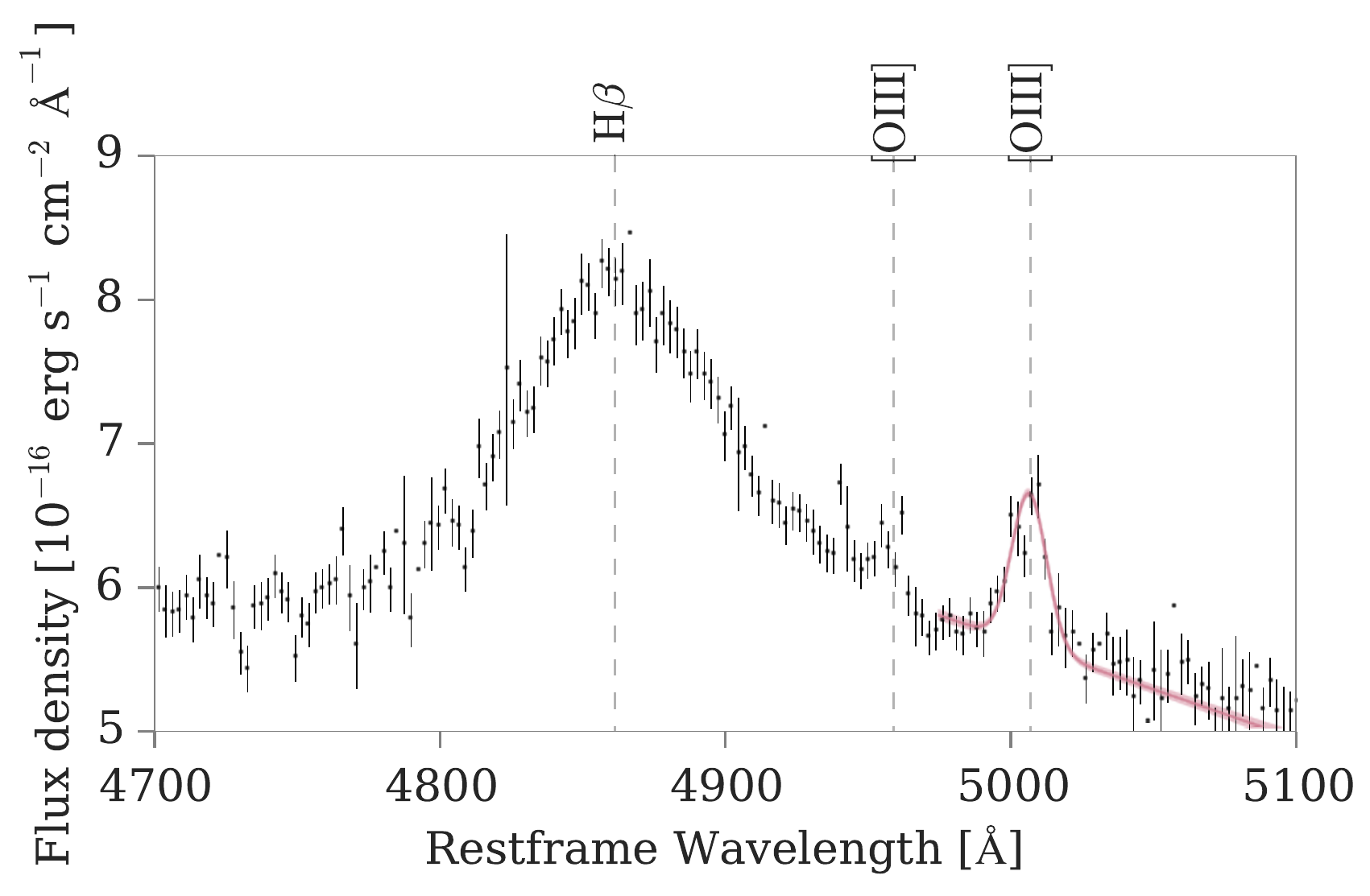}
  \caption[]{Gaussian fit to [OIII]$\lambda$5007 for the object SDSS1354-0013. The red solid line is the
linear least squared best fit. Gray dashed lines indicate the
position of neighboring lines at the redshift of [OIII]. Only every 10th data point is shown for clarity.}
  \figlabel{linefit}
\end{figure}

\subsection{Galactic extinction correction} \sectlabel{extinct}

The spectra are all corrected for Galactic extinction. Extinction
values are queried from the reddening maps of \citet{Schlegel1998}
using the recalibration from \citet{Schlafly2011}. Extinction values
in a 1.5 arcmin radius around the source are extracted from the map yielding four pixels from which an arithmetic average is taken. The reddening law by
\citet{Fitzpatrick1999} with optical total-to-selective extinction
ratio $R_V = 3.1$ is used to deredden each spectrum. The average value
of the reddening for our objects is low, $E(B-V)$ = 0.033, so we
expect any residual effect of the Galactic extinction to be minimal.

\subsection{Resampling} \sectlabel{rebin}

Before combination of the spectra can be done, the spectra are moved
to the local frame. Because of the varying redshifts of the objects,
the spectra will have their spectral arms of X-shooter overlap at
different wavelengths and will therefore have varying
sampling. X-shooter, being a cross-dispersion echelle spectrograph,
has a non-linear dispersion solution and in order to conserve the most
information without oversampling, we need to choose a representative
bin size on which to resample our spectra. The largest sampling is in
the NIR arm and is $0.6$ \AA/pix. An average redshift of $z_{avg}
= 1.6$ gives us the conservative bin size of $0.4$ \AA/pix in the
rest-frame.  
Since our spectra have significantly better resolution than the
nominal values and the UVB and VIS arms have a sampling of $0.2$
\AA/pix we will discard spectral information to make sure that we are
not oversampling. We create a wavelength grid from 1000 -- 11350 \AA~in
steps of 0.4 \AA, giving $\sim$ 16000 spectral elements. The input spectra are moved to the rest-frame
by dividing the wavelengths with $(1 + z)$ and to conserve flux we
multiply the flux density with a corresponding $(1 + z)$. The spectra
are then interpolated using a linear spline and
resampled onto the newly defined common wavelength grid. Since the
constituent spectra have different sampling than the target grid, we
will introduce additional correlations between neighboring
pixels. The high S/N ratio of the individual spectra combined with a goal to
find a spectral slope over a very wide wavelength range implies that the
interpolated pixel-to-pixel uncertainties does not affect the final results.

\subsection{IGM absorption correction} \sectlabel{igm}
At redshifts higher than $z\sim1.6$, the rest-frame coverage of X-shooter
reaches blueward of \lya~and these spectra are therefore affected by
\lya-forest absorption, blueward of the \lya-line. A common way to correct for
this absorption is to obtain a spectrum at a sufficiently high resolution to be
able to trace the continuum emission by visual inspection. At $z < 2 $ the
\lya~forest absorption is not yet heavily blended \citep{DallAglio2008}, and the
X-shooter resolution allows us to effectively trace the continuum. To recover
the continuum we insert points blueward of \lya~where the continuum is visually estimated.
We then manually delete points that are not sufficiently smoothly
varying, signifying points affected by \lya-forest absorption. 
When the points represent the visually estimated continuum, a linear spline is the used to
interpolate between the points onto the sampling of the original wavelength
array. We then concatenate this new continuum estimate with the original
spectrum, at \lya. The objects SDSS1150-0023, SDSS1236-0331 and SDSS1431+0535
are the ones contributing the composite at these wavelengths.

\subsection{Normalisation}  \sectlabel{norm}

Before combination, the spectra needs to be normalized. Depending on the type of features we are interested in and
on the combination method employed, there are different ways to
achieve this normalization. A popular way to normalize is to order the
spectra in increasing wavelength, starting with the lowest redshift,
and scale the entire spectrum by the median value of the flux and in
the overlapping region with the next spectrum. The spectra are then
combined consecutively, always scaling to overlapping region. The
region chosen and the order of combination will affect how the final
spectrum will look
\citep{Francis1991,Brotherton2000,VandenBerk2001,Glikman2006}. We
choose a different normalization method. In principle the variance of
the composite at each pixel should reflect the intrinsic variability of constituent
spectra, but the region to which we scale will affect the absolute
scale of the variance. To make the variance reflect the intra-spectral
variability the most, the spectra have been scaled in the region with
the least intrinsic continuum variability, which decreases with
wavelength, at least up to 6000 \AA~\citep{VandenBerk2004}. The
region closest to 6000 \AA~free of contaminating emission is just
redward of H$\alpha$ in the region 7000 - 7500 \AA. Each spectrum has
been divided by the median flux in this region and therefore the order in which the combination is done does not
matter. Before combination, all spectra have been run
through a filter where if the change between neighboring pixels is greater
than 10 percent, the pixel is flagged as bad and excluded from the
combination. Consecutive comparison is carried out between pixels in order of ascending wavelength. This filtering algorithm is applied to exclude noisy pixels that remain in the spectra after the pipeline processing and pixels affected too
strongly by residuals from the telluric absorption correction. We have investigated
whether this exclusion changes the shape of the combined continuum and
no effect was visible. This way of flagging noisy pixels ensures that
we retain as much signal as possible without introducing very noisy
isolated pixels. As many as 5 percent of the total number of pixels are
rejected in the worst case.

\subsection{Combination}  \sectlabel{combine}

Choosing the right combination method is important and different
methods can yield slightly different results. The first assumption is
that at each pixel of the composite, the constituent spectra have a Gaussian
distribution around a sample mean and that this mean reflects the
expectation value of the sample. Since we have a relatively small
sample, even investigating whether we are sampling from a Gaussian
distribution is ambiguous. To make a test for normality, we take the
mean value of 100 pixels along each on the constituent spectra and, assuming that the continuum is
flat in the region chosen, make a quantile-quantile plot, shown in \Fig{normality}. Given that the constituent spectra are normally distributed, the blue points should
fall on the red line. From
\fig{normality} it is not immediately evident if we are sampling from
a normal distribution, but it cannot be excluded. We run a
Shapiro-Wilk test where the same 100 pixels from the central part of
the spectrum are used. On each individual collection of pixels the test
gives us a p-value for the hypothesis test,
under the null hypothesis that the sample comes from a normal
distribution. With an average p-value over the pixels of $p = 0.55$ we can at
least not reject, that we are sampling from a normal
distribution. Since we have no prior reason to expect that we are
sampling from a Gaussian distribution and that the distribution is
expected to change with wavelength increasing in width as the distance
to the normalization region increases, a certain amount of caution is
warranted.

\begin{figure}[t!]
  \centering
  \includegraphics[width=0.99\columnwidth]{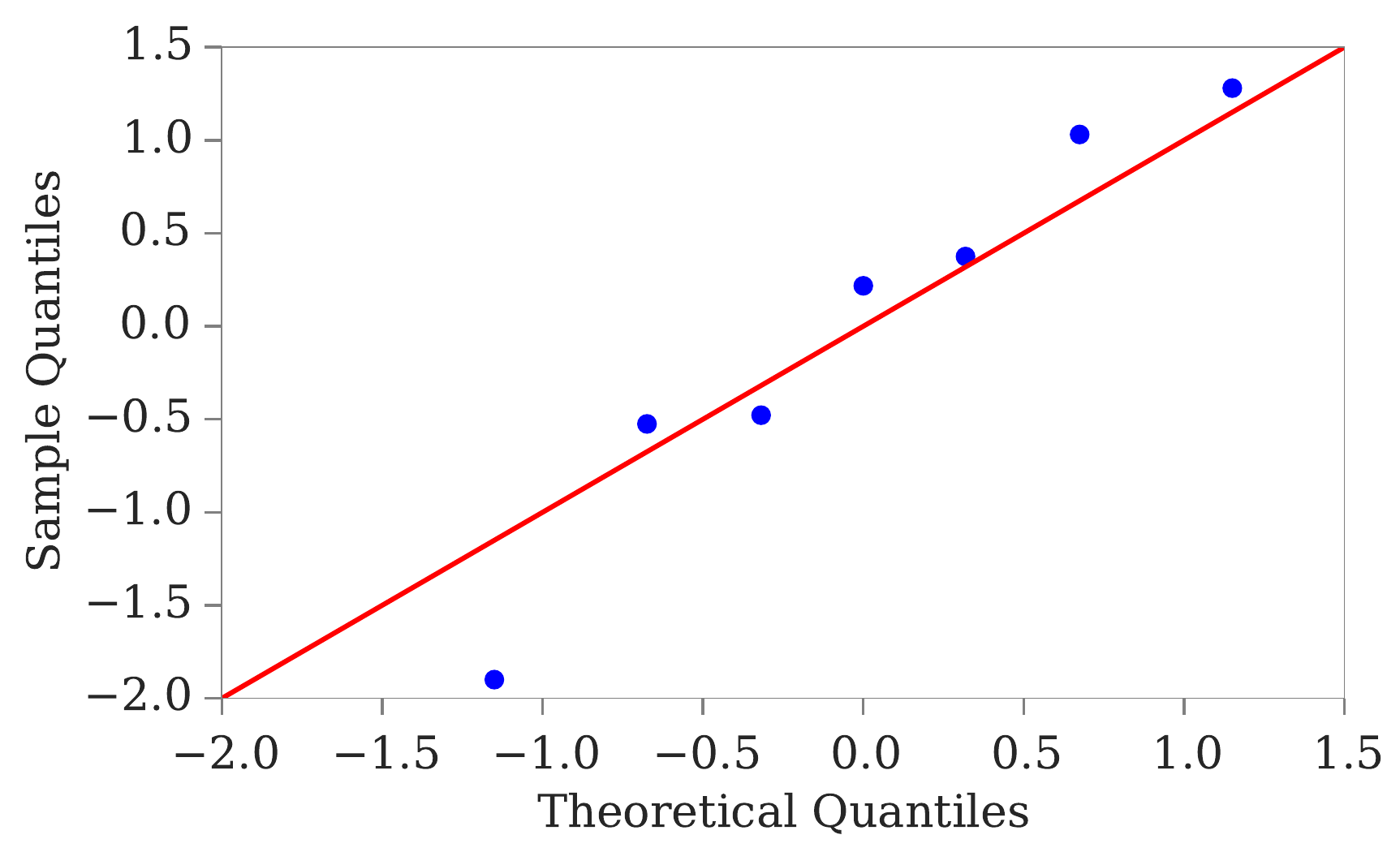}
  \caption[]{Quantile-quantile plot as a test for normality. An average over 100
pixels in the central regions where the continuum is assumed to be relatively
flat is used. The inverse cumulative distribution for the sample is constructed
and plotted against the corresponding quantiles for a normal distribution. The
red line arises if we are sampling for a normal distribution. Since the points
are slightly steeper, this indicates a more dispersed distribution.}
 \figlabel{normality}
\end{figure}

In the case of uncorrelated pixels the minimal variance point estimation for the
sample mean is the inverse variance weighted mean, given again that the
constituent spectra at each wavelength bin can be treated as stochastic
variables that follow a normal distribution $\mathcal{N}(\mu, \sigma_i^2)$ with
mean $\mu$ and variance $\sigma_i^2$. The inverse variance weighted mean is
calculated as
\begin{eqnarray} \eqlabel{wmean}
\bar{f_{\lambda}} &=& \frac{ \sum_{i=1}^n \left( f_{\lambda, i} \sigma_{\lambda,
i}^{-2} \right)}{\sum_{i=1}^n \sigma_{\lambda, i}^{-2}},
\end{eqnarray}
with the variance of the weighted mean given as

\begin{eqnarray} \eqlabel{sigma-wmean}
\sigma_{\bar{f_{\lambda}}}^2 &=& \frac{ 1 }{\sum_{i=1}^n \sigma_{\lambda,
i}^{-2}}.
\end{eqnarray}
Because we have resampled the pixels to a common pixel scale, additional
correlation between adjacent pixels will be introduced both in the flux spectrum
and in the error spectrum, meaning that we will underestimate the errors in the
resulting composite spectrum where the missed statistical noise will be hiding
in pixel-to-pixel variations. Additionally, since our objects are fairly bright,
the Poisson noise in our spectra will be non-negligible and thus we are
incorporating the pixel values themselves into their weights, which leads to a
bias in the result toward spectra of lower signal-to-noise. We check how the
combination method employed affects the qualitative features in our spectrum in
\sect{systematics}.

\subsection{Absolute magnitudes} \sectlabel{absmag}

In order to make a meaningful comparison with other composites and to the parent
population of quasars, synthetic apparent magnitudes are calculated and then,
using the redshift, converted to absolute magnitudes. Synthetic photometry is
the process by which the apparent magnitudes in a bandpass is obtained from a
spectrum by convolution with the filter quantum efficiency curve
\citep{Bessell2005}. Following \citet{Bessell2012} and \citet{Casagrande2014} we
calculate the synthetic AB magnitudes using

\begin{eqnarray}\eqlabel{abmag}
m_{\textmd{AB}_x} &=& -2.5 \log_{10} \left(  \frac{\int f_\lambda \,  T_{x}  \,
\lambda \,  d\lambda} 
{\int  T_{x} \,  (c / \lambda) \,  d\lambda} \right) - 48.60,
\end{eqnarray}
where $f_\lambda (\lambda)$ is the observed flux density in units of \fluxunit~
for the spectra and $T_{x} $ is the total system
fractional throughput in band $x$. The apparent magnitude can then be converted
to absolute magnitude using

\begin{eqnarray}\eqlabel{absmag}
M_{\textmd{AB}_x} &=& -5 \log_{10} \left[  \frac{D_L }{10 \mathrm{pc}}   \right]
+ m_{\textmd{AB}_x} ,
\end{eqnarray}
with the \textit{luminosity distance}, $D_L$,  calculated using a standard
cosmological calculator\footnote{astropy.cosmology}. Using the SDSS i-band filter transmission curve\footnote{\url{http://classic.sdss.org/dr7/instruments/imager/filters/i.dat}}, interpolated to the spectral element size, we get the absolute i-band magnitudes, M$_i(z=0)$.
The largest compilation of quasar absolute magnitudes to date is
\cite{Paris2014}, in which the K-corrected, i-band magnitudes normalized at $z = 2$, $\mathrm{M_i (z=2)}$, for the 166,583 quasars in DR10 are
presented. To get the corresponding magnitudes for the composite constituent
quasars at $z = 2$, we follow the prescription from \cite{Richards2006b} in which the absolute magnitude for an observer at $z = 0$ is converted to that of an observer at $z = 2$. This is done to minimize the systematic effect on the magnitude of the K-correction arising from the varying continuum slopes of quasars. The farther away that a given quasar is from the normalization redshift, the greater the error from a wrongly assumed continuum slope. Choosing $z = 2$ as the normalization redshift, which is close to the peak in quasar number density as a function of time \citep{Richards2006b, Hopkins2007}, minimizes this error. We have not substituted the assumed slope of $\alpha
  = 1.5$ with the updated value, for consistency with the comparison sample.

\section{Results}   \sectlabel{results}

We show the weighted mean composite in \fig{combined}. Characteristic features
of quasars are readily visible. We see several prominent emission lines across
the entire spectral range, which includes both very broad high-ionization lines
and narrower low-ionization lines as is typical of quasar spectra, with both
broad and narrow lines superposed with multiple components \citep{Baldwin1995}.
We have marked the position and name of the most prominent lines and overplotted
them in \fig{combined}.  Analyses of the lines are beyond the scope of this
paper.

 \begin{figure*}[t!]
   \centering
   \includegraphics[width=1.99\columnwidth]{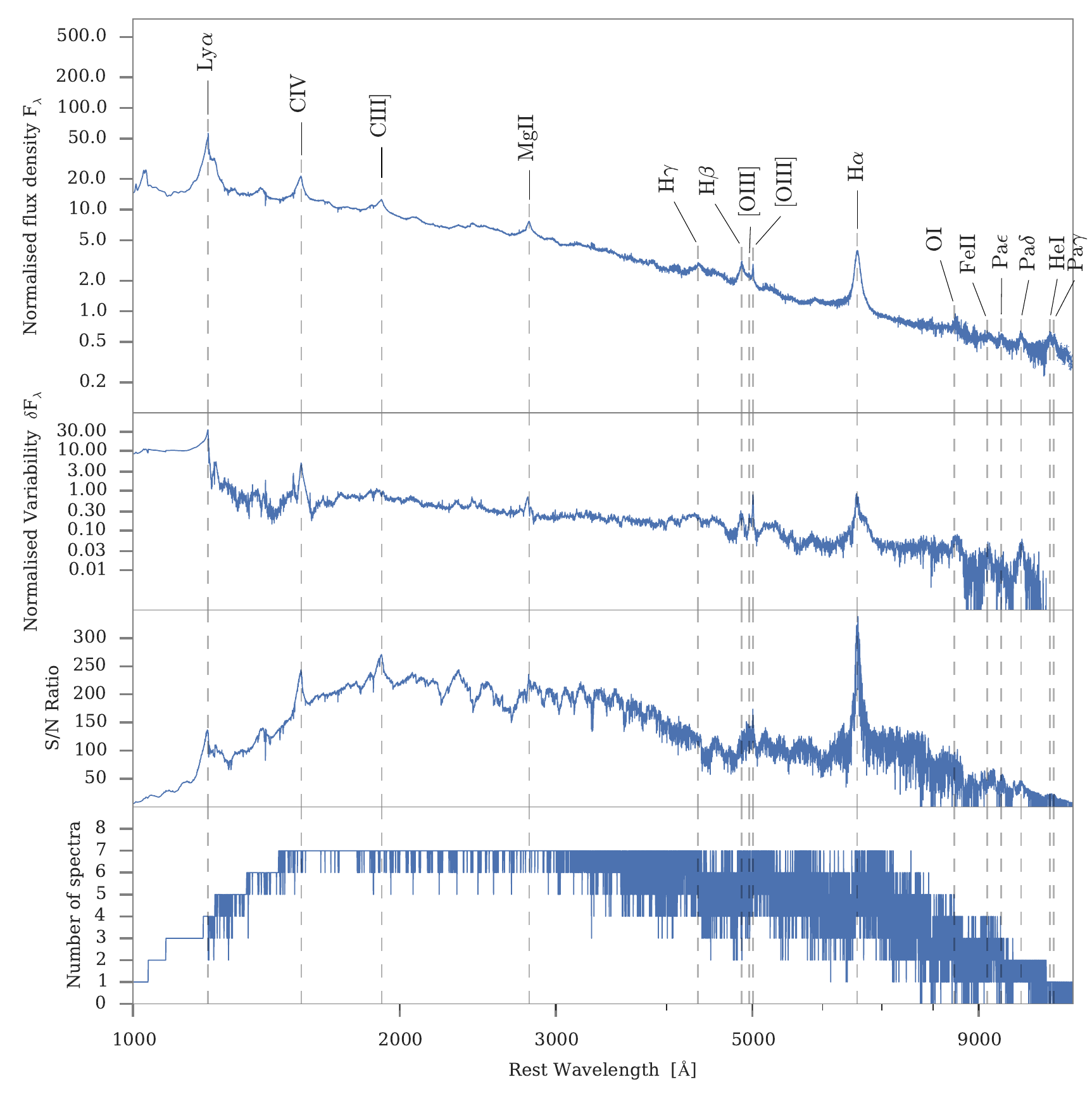}
   \caption[]{\textit{Top panel:} Composite spectrum with prominent emission
lines marked. \textit{Upper middle panel:} Measure of intra-spectrum
variability. The standard deviation of the constituent spectra as a function of
wavelength, where to normalize the differing slopes between the different spectra,
the fitted slope in each of the constituent spectra have been subtracted.
   \textit{Lower middle panel:} Signal-to-noise ratio. The composite spectrum is
divided by the error spectrum, thus directly giving a measure of the signal
strength.
   \textit{Bottom panel:} Number of contributing spectra as a function of
wavelength. 
   
   }
  \figlabel{combined}
 \end{figure*}

As is often seen, the continuum level is shrouded by a myriad of emission lines,
Balmer continuum and Fe-complexes scattered across the entire spectrum, making
the intrinsic shape of the continuum difficult to estimate \citep{Elvis2001}. 
The composite spectrum covers from \lyb ~in the bluest part of our spectrum to
Pa$\gamma$ in the reddest, where a power law adequately describes the
continuum from \lya~and redward. Blueward of \lya, the spectral
continuum changes slope toward a shallower spectral slope. The change in slope
is a consequence of the accretion disk mechanism responsible for the continuum
emission where the position of the spectral break is set by the temperature of the
accretion disk \citep{Pereyra2006}. The correction for \lya-forest absorption is
described in \sect{igm}.
Between $\sim$2000 \AA~and $\sim$5000 \AA, excess emission compared with a
pure power law is detected, consistent with the \textit{small blue bump}
\citep{Wills1985}, which consists of a blend of Balmer continuum and \feii
~lines, with both broad and narrow components.

 The continuum of a single quasar spectrum is usually modeled as a power law, $f_{\lambda}
\propto \lambda^{\alpha_{\lambda}}$, but a weighted mean spectrum of a sample of power-law continua
does not, in general, result in a power law with a slope equal to the mean slope of
the individual quasar spectra. In principle, a geometric mean of a sample of power laws
\textit{should} result in a power law with a spectral index being the mean of the individual spectral indices, as is shown in
\App{math}. We compare the composite obtained by taking the geometric mean with
that obtained by the weighted average and no significant difference is detected
between the two composites. We therefore take the weighted mean composite as
representative of the constituent spectra. This potential cause of systematic
error due to combination methods is investigated more in \sect{systematics}.

Regions free of contaminating line emission are very scarce and fitting a
power law to the composite in manually specified regions is not guaranteed to
reflect the spectral index of the continuum. A quantitative measure of the shape
can be obtained by carefully selecting regions that appear relatively unaffected
by line emission in excess of the continuum. The regions that cover the widest
wavelength range are: 1300 - 1350, 1425 - 1475, 5500 - 5800, 7300 - 7500\AA,
which we use to fit both with a single power law and a broken one, with a
break redward of \hb~at 5000 \AA~as is reported by other authors
\citep[e.g.,][]{VandenBerk2001}. For the single power law we obtain
$\alpha_\lambda = -1.70$, where for a broken power law, we obtain a spectral
index $\alpha_\lambda = -1.69$ below 5000 \AA, and $\alpha_\lambda = -1.73$
above, consistent with a single power law describing the continuum from \lya
~to Pa$\gamma$. The break in \cite{VandenBerk2001} is attributed to
contamination from the host galaxies \citep{Glikman2006} and adapting the same
interpretation, gives support to the assumption about the negligible host galaxy
contribution present in the composite presented here. A detailed comparison with
existing composites is done in \sect{comparison}.

\subsection{Applicability of the composite}  \sectlabel{application}

To test the applicability of the composite, it is used to determine the dust
content of a sample of three red quasars taken from the High A$_V$ QSO (HAQ)
survey \citep{Krogager2015}. The HAQ survey consists of quasars selected on the
basis of their optical colors in SDSS and near-infrared in UKIDDS to identify
the reddest and therefore most dust-extincted quasars which are missed by
traditional color selection criteria. A complete description of the sample
criteria and the data are presented in \citet{Fynbo2013} and
\citet{Krogager2015}. The dust content is inferred by reddening the composite
with an extinction law parameterized by \citet{Gordon2003} to match the object
spectrum. The parameterization allows the redshift of the object quasar and the
magnitude of visual extinction, A$_V$, to be found by minimizing the residuals
between the object and the composite. A detailed description of the fitting
method is given in \cite{Krogager2015}.
The three quasars have been selected to have varying amounts of extinction and
have differing redshifts. We show the result for quasars HAQ2221+0145,
HAQ1115+0333 and HAQ2231+0509 in \fig{application} where the success of matching
the composite with quasars of very different shapes is visible. 
Following \cite{Wang2012}, the previous composite used to fit for the dust
amount consists of the composite generated by \cite{VandenBerk2001} stitched
together with the composite by \cite{Glikman2006} at 3000 \AA. Thus, two
different composites built from differing samples have been treated as a single
composite. That this has been successful is a testament to the similarity of
quasars across a wide range of apparent physical conditions. 
The values obtained with our composite is entirely consistent with the values
published in \cite{Krogager2015}, which is encouraging for the previous use of
the stitched template. A detailed comparison between the different composites is
presented in \sect{comparison}. 

There exists a distribution of slopes for quasar continua and when using the composite presented here, the intrinsic slope of the modeled quasar is highly degenerate with the amount of dust inferred \citep{Reichard2003a}. Since attenuation by dust is stronger at shorter wavelengths, extinction introduces a curvature in the spectrum, which is separate from a pure change in slope, see
\citet{Krawczyk2015} for a discussion of \textit{red vs. reddened} quasars.. 

\begin{figure}[t!]
 \centering
 \includegraphics[width=0.99\columnwidth]{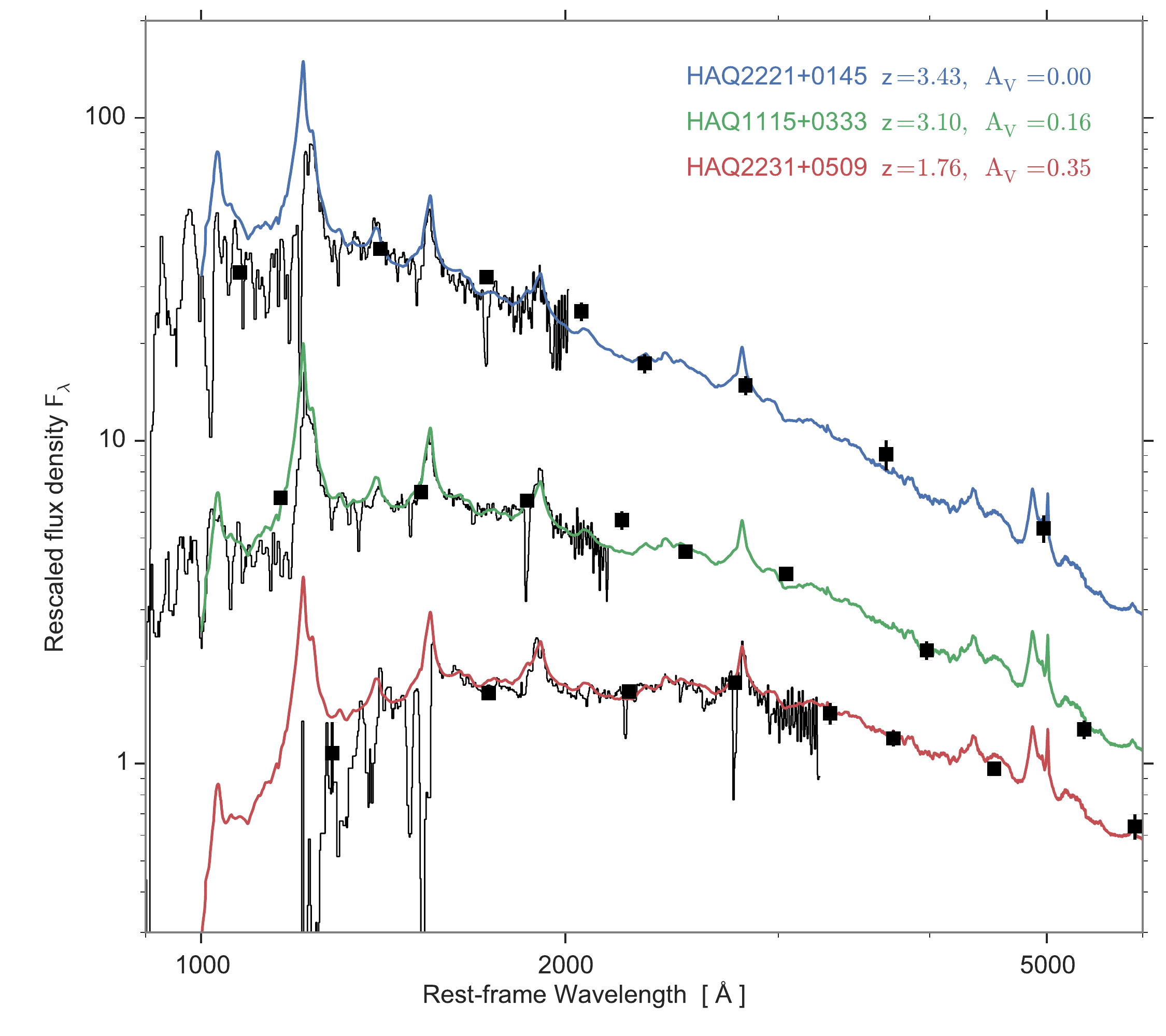}
 \caption[]{Dust extinction, $A_V$, and redshift has been inferred
   for three HAQ quasars by fitting for the amount of reddening
   required to match the template with the observed spectrum. The
   black spectra are taken at the Nordic Optical Telescope (NOT) using
   the ALFOSC spectrograph. The black squares are optical $ugriz$
   photometric points from SDSS and UKIDSS $YJHK$
   bands. For varying amount of extinction at different redshifts,
   consistent results are found with previous usage of stitched
   composites.}  \figlabel{application}
\end{figure}
 
 \subsection{Intra-spectrum variability}  \sectlabel{variability}
 
To gauge the intra-spectrum variability, the power-law slope of each individual
spectrum is subtracted from the flux points at each wavelength thereby removing
the variability due to different slopes. The standard deviation is then taken of
the normalized spectra, which then reflects the variability of the spectra around
the composite. We show the computed
 normalized variability in \fig{combined}. The normalized sample variability
contains both the temporal variability and the intrinsic population variability,
which is not possible to separate with single-epoch data. 
Significant variation is visible in the lines of \civ,
\mgii, \oiii~and lines of hydrogen as well as in the \feii-complexes
centered on \hb. From \civ~to \hb~an increase in continuum
variability is observed, likely due to the quasar-to-quasar
variability of the small blue bump. In the region blueward of \lya~the fractional variability reaches 60 per cent, attributed to the
stochastic nature of the Lyman-$\alpha$ forest.

\section{Discussion}  \sectlabel{discuss}

The applicability of quasar composites as tools for template matching
and dust estimation largely hinges on the ability of the composite to
represent an intrinsic target spectrum. The broad usage and success of
previous composites largely confirms this ability. That the
constructed composites are representative of a large group of
differently selected quasars is a testament to the homogeneity of the
quasar population and the universality of the emission mechanism.  The
remarkable uniformity of the average spectral properties across
luminosity and redshift indicates very similar underlying physical
mechanisms that can be understood in terms of Eigenvector 1
\citep{Boroson1992, Francis1992} where the Eddington ratio drives the
relative strength of the lines, and orientation effects influences the
observed kinematics of the lines \citep{Shen2014a}. The specific local
physical conditions that regulate the accretion rate then account for
the majority of the inter-quasar variation. In this section, we will
investigate whether the targets selected are simply scaled-up version
of average quasars, or remarkable in some sense.

\subsection{Comparison to global population}  \sectlabel{parents}

The QSOs selected for this study lie amongst the brightest values of
M$_i(z=2)$ compared to the spectroscopically confirmed SDSS QSOs
\citep{Shen2011}, which are uniformly selected to $\sim$90 percent
completeness \citep{Richards2002, VandenBerk2005}. It is clear that
the sample presented here represents some of the brightest quasars
existing between redshift 1 and 2. It is therefore potentially a
biased subset of the global quasar population. Quasars have previously
been shown to exhibit a high degree of homogeneity over many orders
of magnitude \citep{Dietrich2002}, but this is only true for quasars
resembling \textit{average} quasars and not exotic sub-types. To
ensure that the targets selected for this sample are not outliers in
color, the spectrophotometric colors are compared to those of the
quasar population presented in
\citet{Paris2014}\footnote{\url{https://www.sdss3.org/dr10/algorithms/qso_catalog.php}}.
We
show the comparison in \fig{color_comparison}. From the magnitudes, we
first confirm that the quasars selected here are among the
intrinsically brightest visible in $i$-band with a 2.7 sigma magnitude
distance from the mean absolute $i$-band magnitude, normalized at redshift 2,
M$_i(z=2)$ of the full SDSS quasar catalog for DR10. Despite their luminosities the colors are representative
of the subset fulfilling the selection criteria. As a consistency
check, the composite generated in this work is used to fit for the
dust content of quasars with varying degree of extinction and
consistent results were obtained as compared with composites generated
from other subsamples of the parent population. For a comparison
between SDSS and quasars detected in \textit{GALEX}, UKIDSS,
\textit{WISE}, 2MASS and \textit{Spitzer}, see
\citet{Krawczyk2013}.

 \begin{figure}[t!]
   \centering
   \includegraphics[width=0.99\columnwidth]{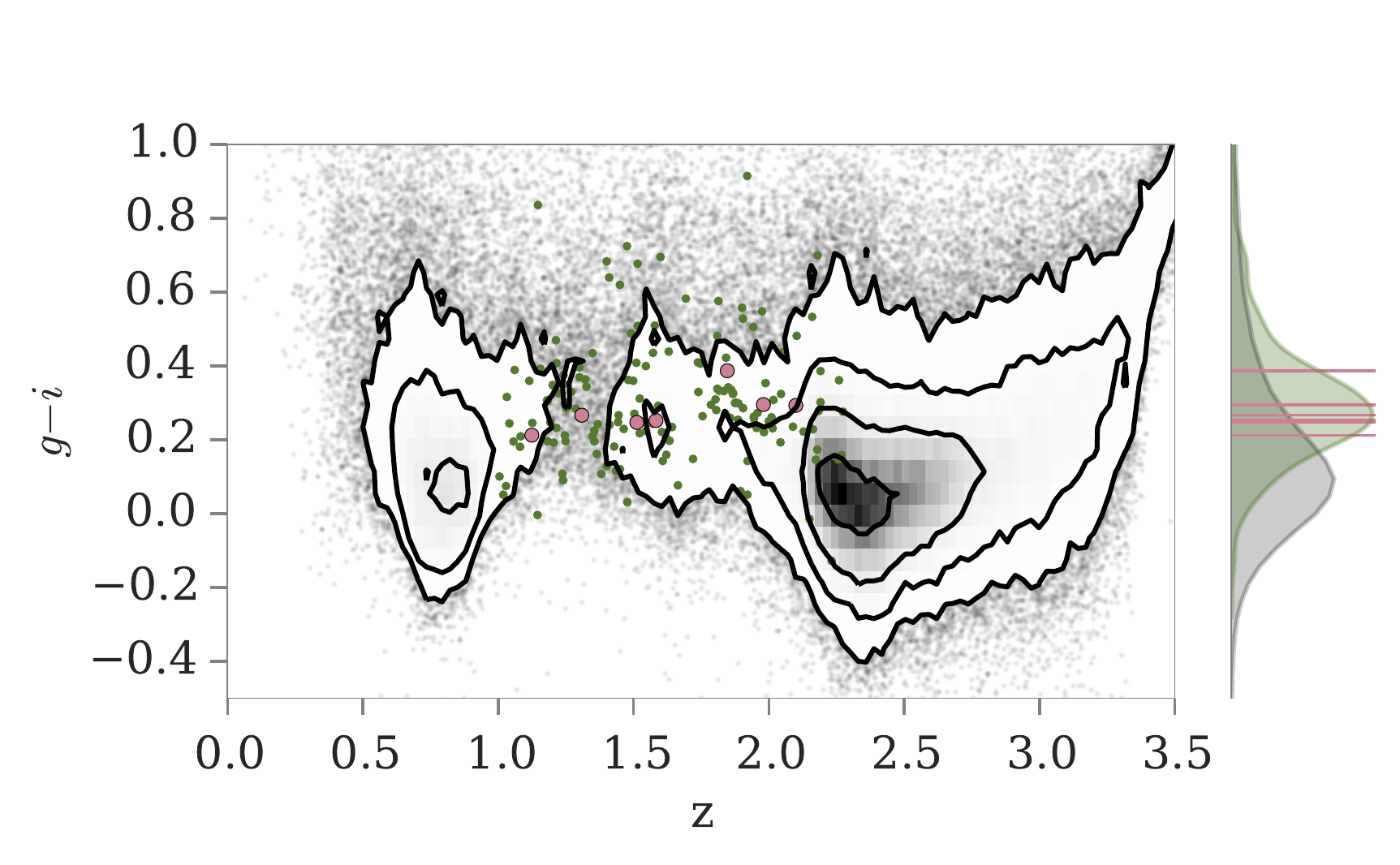}
   \caption[]{Quasar color, \textit{g-i},  as a function of redshift, z. The
full quasar sample from SDSS DR10 \citep{Paris2014} is shown in gray color.
Bounding contours show the 0.5, 1, 1.5, 2 $\sigma$ contours of the density of
points. The olive points are individual quasars from the SDSS sample which
satisfy $r <= 17$. Overplotted in pink are the quasars contributing to the
composite presented here. The quasar colors have been marginalized over in the
right side where the tall pink lines are at the position of the composite
quasars and the olive shaded area is the kernel density estimation of the
distribution of quasars fulfilling the selection criteria. The gray-shaded area
is the projection of the DR10 quasar sample.}
  \figlabel{color_comparison}
 \end{figure}

Because of the high luminosity of the objects presented here, several
differences in the properties are expected as compared to those of the
global quasar population. The physical scales locally are expected to
be larger \citep{Bentz2013}, which will make variability timescales
longer and variability magnitudes smaller \citep{VandenBerk2004,
  Schmidt2012}. Black-hole masses as determined from \mgii~increases
with luminosity \citep{wu2015} which in turn affects the temperature
of the accretion disc \citep{shakura1973, Pereyra2006} and thereby the
position of the "big blue bump" and the degree to which the continuum
is well modeled by a single power law \citep[see also][for a
  discussion]{Lusso2015}. As can be seen from \fig{color_comparison},
the quasars selected for the composite represent average colors for
the subset fulfilling the sample cuts. An immediate consequence of
this will be that the amount of dust inferred using this composite
will be an average content over a statistical sample, because the
individual object which is sought modeled, can have an intrinsic
quasar continuum shape that is different from what is constructed here
\citep{Richards2003, Hopkins2004}. Populations with intrinsically
different spectral slopes exists \citep{Glikman2012, Krawczyk2015} and
the bias in the inferred amount of extinction is therefore larger in
these types of surveys. Since the \textit{true} intrinsic dereddened
quasar color distribution is likely broad with a range of slopes, a
caveat of using this composite is that the amount of extinction
inferred will be the average amount over a large sample.

\subsection{Comparison to the sample population}  \sectlabel{sample_pop}

\begin{figure*}[t!]
\centering
\includegraphics[width=1.99\columnwidth]{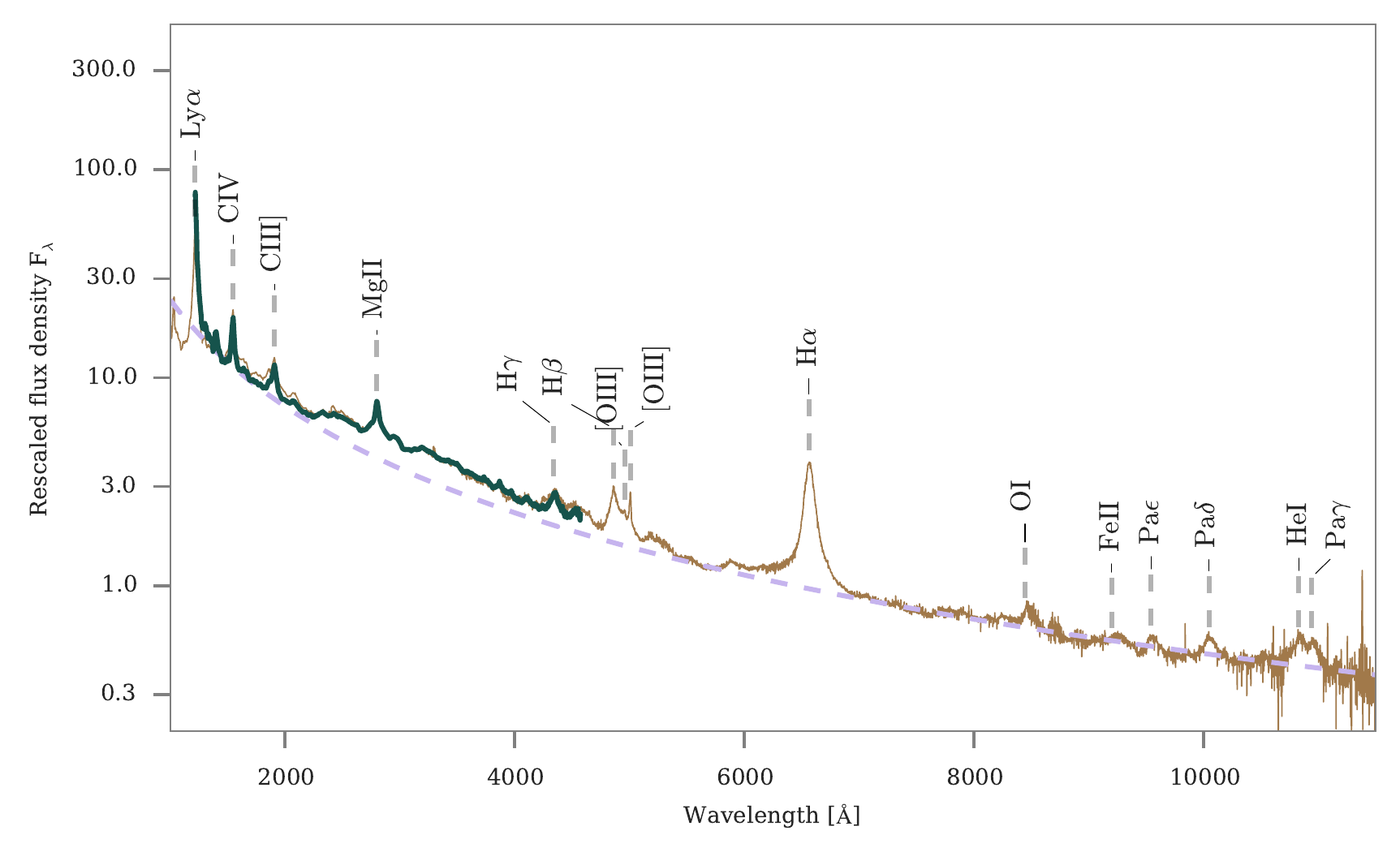}
\caption[]{X-shooter weighted arithmetic mean quasar composite on a linear wavelength
scale in light brown. The positions of several prominent emission lines are
marked. Overplotted in dark green is the corresponding composite generated from
the full sample of SDSS quasars fulfilling the selection criteria and general
agreement is observed, albeit with a brighter Balmer continuum in the
SDSS-constructed composite. In purple is shown the results from fitting both a
pure and a broken power law to the regions specified in \sect{results} and they
are observed to be indistinguishable.}
\figlabel{composite}
\end{figure*}

Because of the low number statistics used for the sample presented in this work, a
further bias can be that the targets are not representative of the
population fulfilling the cuts imposed in the selection. As
highlighted in \fig{color_comparison}, the objects selected here are comparable in color to the full sample fulfilling the selection criteria. To
quantify the bias against the selection, we construct a composite in
an identical fashion consisting of all SDSS quasars fulfilling the
selection criteria.  In \fig{composite} we overplot the composite
generated from the SDSS spectra of the full sample fulfilling the
selection criteria on the one constructed with X-shooter. It is clear
that there are small differences in the shape of the composite
constructed from the full sample and the small sample, but
qualitatively the shape remains the same. Calculating the power-law
slope in each individual SDSS spectrum yields $\alpha_\lambda = -1.6\pm 0.3$,
but this number is difficult to interpret due to the narrow continuum
region free of emission lines (1300 - 1350 \AA~and 1425 - 1475
\AA). It is reassuring nonetheless that they are consistent within the
errors as seen in \tab{comparison}. We investigate the effect of a
finite sample size on the combination method in \sect{Combination method}.

\subsection{Comparison to existing composites} \sectlabel{comparison}
We compare the X-shooter composite to composites from the literature by the following authors:
\citet{Francis1991, VandenBerk2001, Telfer2002, Glikman2006} and
\citet{Lusso2015}. \citet{Francis1991} generated a composite from 688
LBQS (Large Bright Quasar Survey)-quasars observed mainly with the Multiple Mirror Telescope (MMT). The absolute magnitudes cover $\sim -22$
and $\sim -28$ at a redshift range $0.05 < z < 3.36$ subject to a
significant Malmquist bias for which the mean absolute magnitude is a
strong function of rest-frame wavelength, with the brightest objects
contributing at shortest wavelengths and visa versa for the faint
objects. The composite by \citet{VandenBerk2001} consists of 2204 SDSS
quasars at $z_{median} = 1.253$ with absolute M$_{r'}$-magnitudes
between $-18.0$ and $-26.5$ and therefore intrinsically fainter and
lower redshift objects. \citet{Lusso2015} calculated the M$_i(z=2)$
magnitudes for the 184 constituent objects in \citet{Telfer2002}
observed with FOS, GHRS and STIS onboard the Hubble Space Telescope and find an average
M$_i(z=2) \sim -27.5$ at $z \sim 1.2$, which yields a composite of
fainter, more nearby sources. For construction of a composite,
\citet{Glikman2006} used 27 objects at $z \sim 0.25$ with an average
absolute \textit{i}-band magnitude M$_i \sim -24$ observed with IRTF,
therefore constituting a fainter, lower redshift sample. Lastly,
\citet{Lusso2015} used 53 quasars observed with WFC3 at $z \sim 2.4$
with absolute i-band magnitudes at redshift 2, M$_i(z=2) ~\sim -28.5$,
so of comparable brightness, but at slightly larger distances.

Regardless of the variation across selection criteria, luminosities and
redshifts, a remarkable similarity in the overall shape is visible
over a wide wavelength range. The simultaneous existence of lines
arising in such different environments around objects differing by
many orders of magnitude in luminosity is caused by the varying
conditions in the emitting clouds which ensure that each ionic species has optimal conditions to produce line emission \citep{Baldwin1995}.
Significant differences between the composites are visible blueward of
\lya~where different methods for IGM absorption correction has been
employed. No correction has been applied in the composites by
\citet{Francis1991} and \citet{VandenBerk2001} and a sharp decrease in
continuum flux is detected. Because higher redshift objects contribute
mostly at shorter wavelengths, significant \lya~absorption is
expected regardless due to the \lya~opacity evolution
\citep{Moller1990, Madau1995, DallAglio2008}. The IGM correction of
\citet{Telfer2002} is a hybrid in which some manual and some
statistical correction has been applied. \citet{Lusso2015} rely on a purely statistical correction that is calculated
by integrating the neutral hydrogen column density distribution at $z
\sim 2.4$ measured by \cite{Prochaska2014b}. The IGM correction in the
X-shooter composite presented here is done in each individual spectrum
where resolved lines are interpolated over. Good agreement is found in
the \lya~forest region with \cite{Telfer2002}, but underestimated as
compared to \cite{Lusso2015}, which can be seen in the insert in
\fig{composite_comparison}. As argued by \cite{Lusso2015}, an
explanation for this discrepancy is the underestimation of the
correction needed to account for unidentified Lyman limit
absorbers. Due to the resolution of the instrument and the redshifts
probed, this should not be a significant effect on the sources
observed here.  Differences in the emission lines are also
visible. This is to be expected due to the highly varying magnitudes
of the constituent objects for the different composites where a
decreasing equivalent line width is expected with increasing
luminosity \citep{Baldwin1977}.  Above 5000 \AA, a clear discrepancy
is visible to the \citet{VandenBerk2001} composite which is due to
host galaxy contamination \citep{Glikman2006}. From the quasars in
DR7 \citep{Shen2011}, it is confirmed that this effect is not
affecting the X-shooter composite presented here to a significant
degree, due to the extreme luminosities. This is investigated further
in \sect{Host galaxy
contamination}.

 \begin{figure*}[t!]
   \centering
   \includegraphics[width=1.99\columnwidth]{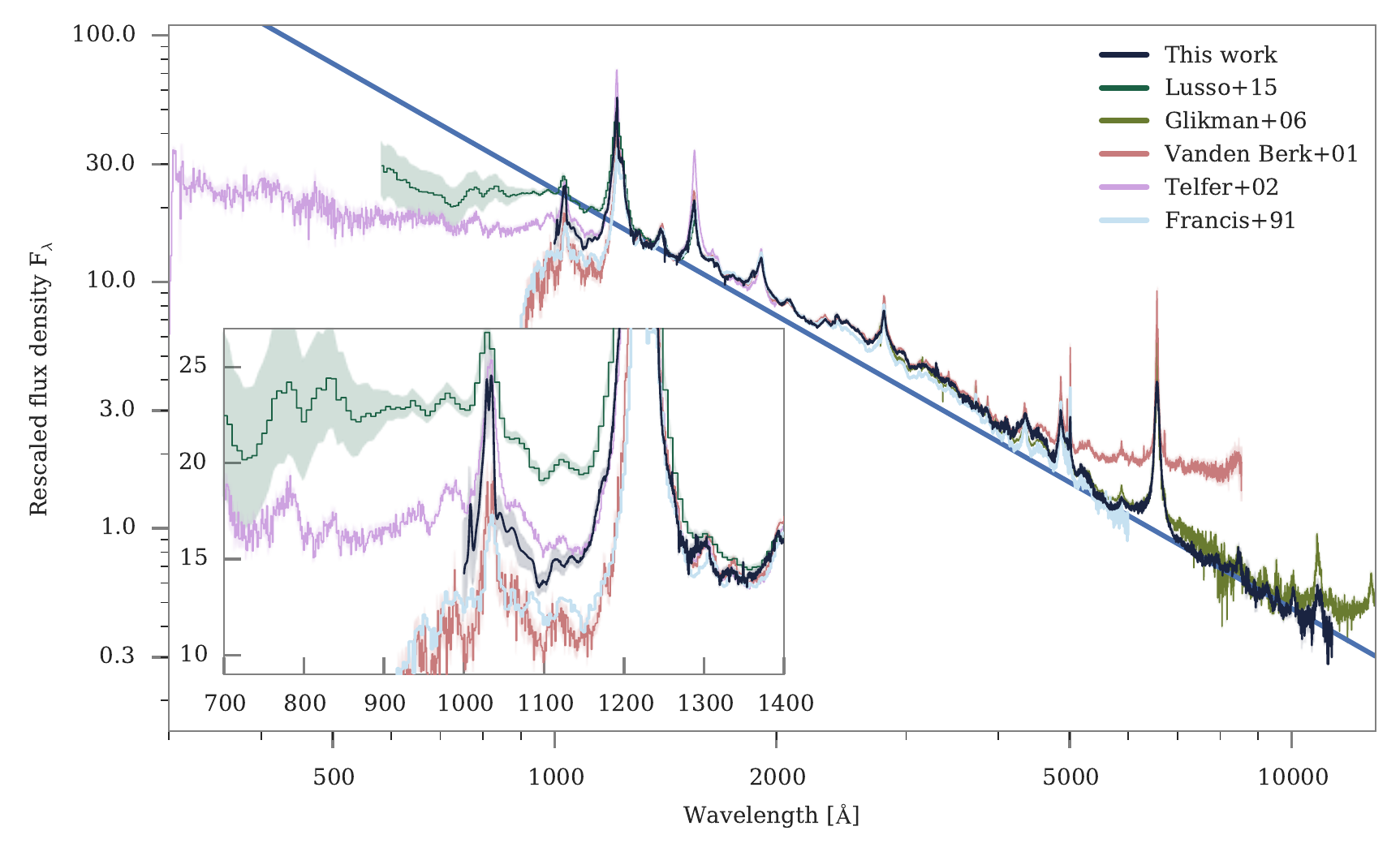}
   \caption[]{Comparison of different composites. The composites by
     \citet{Lusso2015, VandenBerk2001, Telfer2002, Francis1991} are
     normalized to the X-shooter composite at $\sim 1450$ \AA~and the
     composite by \citet{Glikman2006} is normalized to ours at $\sim
     3850$ \AA. Significant differences are visible blueward of
     \lya~due to differing IGM correction methods. Above 5000 \AA~
     significant host galaxy contamination is visible in the composite
     by \citet{VandenBerk2001}. Overplot in blue is a power law
     with slope $\alpha_\lambda = -1.70$ and normalized at $\sim 1450$ \AA.}
   \figlabel{composite_comparison}
 \end{figure*}
 \input{tabs/comparison.tex}

Despite the similarity visible in the direct comparison, it is
remarkable that the power-law slope blueward of 10000 \AA~is found to
be varying by more than 30 per cent between the different authors. We
compile the results from the different composites in
\tab{comparison}. One likely reason for the discrepancies is the
wavelength regions used for the continuum where an increased
wavelength coverage allows a better continuum to be selected,
especially selecting regions unaffected by Balmer continuum and broad
\feii~complexes. A more direct comparison of the slopes, where an
explicit fit to the same regions is made, is not possible due to the
different wavelength coverage. A narrow region free of line emission just redward of \lya~is covered by most of the composites in the comparison, but fitting a power law to such a narrow region for a small sample would largely leave the inferred slope dependent on the exact ranges chosen for the fit.

\subsection{Systematics}  \sectlabel{systematics}

Understanding how systematic effects may affect results is important in making
robust conclusions.
 We list the possible systematics that effect the composite and address each of
the points separately to investigate to what degree they affect the final result
and their potential consequences.

\subsubsection{Combination method}  \sectlabel{Combination method}
What is sought illustrated with the combination method employed is the central
tendency of the selected sample normalized at an optimally chosen region. Since
the underlying distribution is unknown, choosing the optimal point estimator for
the average flux value is ambiguous. The composite presented in \sect{results}
used the weighted arithmetic mean, which was chosen because it is the maximum
likelihood estimator of the mean, given that we are sampling from a normal
distribution with independent sampling. This combination method, however, has
systematic differences from other methods, especially the danger of biasing the
composite toward the highest signal-to-noise spectra. 

To test whether we are biased toward high S/N spectra we compare the
weighted mean spectrum with the composite created from the median. We
perform the comparison by constructing a fractional difference
spectrum, which is a ratio between the two composites. There is very
good agreement between the two composites with a mean fractional
difference of $0.01 \pm 0.05$ per cent. There are no observed trends with wavelength.

\input{tabs/combinations.tex} The systematic error in the reported
slope due to the different combination methods is assessed by fitting
a single power law in the same regions as in \sect{results} for each
combination method and in the constituent spectra. The value of the
spectral indices, $\alpha_\lambda$ are shown in \tab{comb_meth} for the
different composites. The value of and error on the slope is obtained
by resampling the spectra within the errors and refitting a power law for
each realization. The errors used for resampling are the standard
deviations of the constituent spectra at each spectral bin and, where
available, the statistical errors are added in quadrature.
 A weighted least squares fit is performed on the 10000 realizations,
 and the best-fit value is taken as the mean of the fitted slopes. The
 reported error is the standard deviation of the sample of inferred slopes.
 The error on the composites should approach the one reported for
 $\alpha_\lambda$ measured on the individual spectra which reflect the
 intrinsic variability of slopes. The standard deviation of the
 individual slopes is more reasonable and all combination methods are
 consistent within these errors. The values for the individual mean
 and individual median differ by a small amount and this could suggest
 a slight skew in the distribution of slopes, but the low level of
 discrepancy could also be an artifact of low number statistics. The
 fact that the fitted slopes are consistent supports the use of the
 weighted mean composite.

In general the arithmetic mean spectrum is not guaranteed to have a
slope with the mean spectral index $\langle a_\lambda\rangle$ of the
individual spectra, see \app{math}, but what we see here is a very
good agreement. To obtain a spectrum \textit{with} the mean slope of
the individual spectra, the geometric mean \textit{does} give rise to
a slope equaling the mean slope. The low level of discrepancy between
the different methods is encouraging to the use of the weighted mean.

\subsubsection{Selection effects}  \sectlabel{Selection effects}
Bright quasars with ($r \lesssim 17$) at redshifts $1 < z < 2.1$ are
not necessarily representative of the intrinsic quasar population
\citep{Paris2014} and the final composite is not guaranteed to reflect
the properties of quasars at different magnitudes or
redshifts. Comparison to the parent population yields, in terms of
colors, a redder than average selection for the quasars fulfilling the
selection criteria for the X-shooter composite. The objects that
fulfill the selection criteria are $\sim 0.2$ mags redder than the
quasar mean \textit{g-i} color. This translates into changing a pure
power-law slope from $-1.6$ to $-1.9$, so toward steeper slope and
more blue spectra. Since quasar spectra are more complicated than pure
power laws, a more likely explanation for this offset between the
full quasar sample and the subset selected here is the shifting of
\lya-line emission into the SDSS \textit{g}-band, therefore artificially making
the continuum appear bluer.  A comparison to the composite by
\citet{VandenBerk2001} which consists of lower redshift, intrinsically
fainter quasars, shows a high degree of similarity in the general
shape and small differences on small scales, see \sect{comparison}.
\subsubsection{Flux calibration}  \sectlabel{Flux calibration}
As stated in \sect{sample}, we find slight variations in both the
absolute flux level and spectral slope of the X-Shooter observed
spectra compared with those observed with SDSS. Since we are to a
high degree seeing limited, we expect the slit loss to be
insignificant and potential alternative solutions are investigated. To
quantify this effect of variation of the accuracy of our
flux calibration compared with the corresponding object spectra
obtained with SDSS, we compare the photometry obtained from SDSS with
the synthetic magnitudes calculated in \sect{absmag}. We calculate the
mean difference between the two magnitudes. There is a clear trend with highest
deviations in the
\textit{u}-band and a monotonic decrease in deviations toward
\textit{z}-band, with $-0.4 \pm 0.3$ mags in the \textit{u}-band to $-0.1
\pm 0.2$ mags in the \textit{z}-band. This variation is
caused by both the accuracy of the flux calibration and the intrinsic
quasar variability. Since the slit loss is expected to be wavelength
dependent, peaking in the blue, this is consistent with the trend
observed. For the intrinsic variability, \cite{MacLeod2012} find a
characteristic variation timescale of $\sim 2$ years with an average
rest-frame variation of $\sim 0.26$ mag - comparable to what is observed here.
Recent work
by \cite{Morganson2014} confirm the results of \cite{Helfand2001} that
quasars are more variable in bluer bands - also consistent to what is
found here. Since the mean temporal deviation should average to zero and we find
negative mean differences, part of the discrepancy is likely attributed to flux
calibration differences. It is difficult to attribute the total variation to
either explanation.

\subsubsection{Normalization region}  \sectlabel{Normalisation region}
Depending on the region selected for normalization of the individual
spectra, the distribution of constituent fluxes at any given
wavelength changes. This will mostly affect the absolute scale of the
composite and the reported inter-spectrum variability as a function of
wavelength, but will also cause the value of the composite at a given
wavelength to change depending on the choice of central tendency
estimator. To investigate how the normalization-window chosen affects
the shape of the composite, we have generated composites for varying
position of the normalization-window across the entire wavelength
range. We have chosen regions clear of strong emission lines. The
different combination methods yield slightly varying composites as the
relative distribution of the fluxes changes for a changing
normalization region. After rescaling the composites to unity at 6850
\AA~we see very good agreement between the different normalization
regions.

 We fit a power-law slope to composites generated with varying normalization
regions and take the standard deviation of the fitted slopes as the 1-$\sigma$
systematic error due to different normalization regions and find a value of:
$\sigma_{n} = \pm 0.03$.

\subsubsection{Telluric correction}  \sectlabel{Telluric correction}
To test the magnitude of the noise introduced by the telluric correction we
construct a composite where instead of correcting for the atmospheric absorption
we exclude regions of low transmission, namely the two absorption band at $\sim
14000$ and $\sim 19000$\AA. The effect on the composite is largely an increase
in noise in the regions where individual spectra have been masked, but more
problematic is the introduction of residual atmospheric absorption in unmasked
regions. Since there is significant absorption from $\sim 7000$ \AA~and
redward, as is also visible in \fig{telluric_qc}, and especially in the NIR
arm, masking each individual absorption complex throughout the spectrum is not
desirable. The telluric correction method essentially leaves the unaffected
region free of change.

 \subsubsection{Host galaxy contamination.}  \sectlabel{Host galaxy
contamination}

A geometric composite has been generated in bins of luminosity by
\citet{Shen2011} and the evolution of the relative contribution from the host
galaxy with luminosity is investigated. It can be seen that the lower luminosity
quasars have a spectral break around $\sim$ 4000 \AA, which is lacking at higher
luminosities and attributed to host galaxy contamination. We calculate
$L_{5100}$ for the spectra in the composite presented here, using the continuum
flux density at $5100$ {\AA} and converting to luminosity following
\cite{Netzer2007},
\begin{eqnarray}\eqlabel{l5100}
L_{5100} &=&    4 \pi D_{L} ^{2} \lambda_{5100}  F_{\lambda5100},
\end{eqnarray}
we find and average $\log L_{5100} = 46.7 \pm 0.1$ which is an order of
magnitude higher than the highest luminosity bin in \cite{Shen2011} which is
assumed to be free of host galaxy contamination. This is consistent with what is
found by \citet{Hopkins2007}, that host galaxy light is likely a small
contribution for quasars brighter than $M_{B} \gtrsim -23$. We have selected our
quasars to be very bright and at high redshift thereby reducing the spectral
contamination from underlying host galaxy light.

\section{Conclusion}  \sectlabel{conclusion}

We have generated a quasar composite covering the entire range from 1000 \AA~to
11350 \AA~ based on observations with X-shooter of seven bright ($r \lesssim 17$)
quasars at redshifts $1 < z < 2.1$ free of host galaxy contamination. Assuming a
power-law continuum, we found a spectral slope of $\alpha_\lambda = -1.70 \pm 0.01$, which is
somewhat steeper than for composites presented by other authors. We
found a consistent slope within errors for various combination methods. We
attributed the discrepancy between the power-law slopes to the different wavelength
coverages, where the very wide wavelength coverage of X-shooter ensures that we
can effectively choose regions free of emission lines and line-continuum
emission, especially the Balmer continuum and \feii-line complexes, i.e., the
Small Blue Bump heavily affecting the region from $2000 - 5000$ \AA. We
applied the composite to fit for the amount of dust in three quasars with
varying dust content and found that the amount of visual extinction 
agrees with what is found using a combination of previously generated
composites. A comparison with other composites was done, and a high degree of similarity is highlighted despite the very different selection criteria and
reported spectral slopes.
We have made the composite and all the code used to generate it available at
\url{https://github.com/jselsing/QuasarComposite}. We show a portion of the
composite values for clarity regarding the format in \tab{example}.

\input{tabs/example.tex}

\section{Acknowledgements}  \sectlabel{Acknowledgements}

We thank the anonymous referee for a constructive report that improved our
manuscript on several important points. We thank Marianne Vestergaard and Jens
Hjorth for useful discussions and suggestions. The research leading to these
results received funding from the European Research Council under the
European Union’s Seventh Framework Program (FP7/2007-2013)/ERC Grant agreement
no. EGGS- 278202. LC acknowledges support from an YDUN grant DFF – 4090-00079.
The Dark Cosmology Centre was funded by the DNRF. This research made use of
Astropy, a community- developed core Python package for Astronomy
\citep{TheAstropyCollaboration2013}. The analysis and plotting has been achieved
using the Python-based packages Matplotlib \citep{Hunter2007}, Numpy, and Scipy
\citep{VanderWalt2011}, along with other community-developed packages.

\bibliographystyle{aa_arxiv}
\bibliography{references}


\appendix

\section{Mean spectral index}  \sectlabel{math} \label{math}

We derive that the geometric mean of a sample of power laws equals a power law
with the sample mean spectral index. The geometric mean is a type of mean
defined as
\begin{eqnarray}\eqlabel{geometric mean}
\bar{f_{\lambda}} &=&  \left( \prod_{i=1}^n f_{\lambda, i} \right) ^{1/n},
\end{eqnarray}
where the product is over the individual spectra. We model each spectrum as a
power law
\begin{eqnarray}\eqlabel{powerlaw}
f_{\lambda, i} &=&  k \lambda ^{\alpha_{i}},
\end{eqnarray}
and we get by inserting \eq{powerlaw} into \eq{geometric mean}
\begin{eqnarray}\eqlabel{deriv1}
\bar{f_{\lambda}} &=&  \left( \prod_{i=1}^n k \lambda ^{\alpha_{i}}\right)
^{1/n}.
\end{eqnarray}
We rewrite this expression and make the product a sum in the exponent
 \begin{eqnarray}\eqlabel{deriv2}
 \bar{f_{\lambda}} &=&  k \left( \lambda ^{ \frac{1}{n} \sum_{i=1}^n \alpha_{i} 
}\right),
 \end{eqnarray}
where we see that the geometric mean of a sample of power laws is a power law,
 \begin{eqnarray}\eqlabel{deriv3}
 \bar{f_{\lambda}} &=&  k \lambda ^{ \bar{\alpha_{i} }},
 \end{eqnarray}
with the mean index
 \begin{eqnarray}\eqlabel{mean}
 \bar{\alpha} &=&  \frac{1}{n} \sum_{i=1}^n \alpha_{i} .
 \end{eqnarray}

\end{document}

%% file: tabs/targets.tex
\begin{table*}
\centering
\begin{center}
\caption{Quasars in the composite.}
 \tablabel{targs}
\begin{tabular}{cccccccc}
\hline
\noalign{\smallskip}
SDSS identifier & $\alpha$(J2000) & $\delta$(J2000) & r &  $z_{SDSS}$ $^{(a)}$  &  $z_{fit}$ $^{(b)}$ & $\mathrm{M_i (z=0)}$ & $\mathrm{M_i (z=2)}$\\  
\hline

SDSS0820+1306  & 08 20 45.39 & $+$13 06 18.99 & 15.91 & 1.1257 $\pm$ 0.0001 & 1.1242 $\pm$ 0.0002  &   $-$28.51 &   $-$28.46       \\
SDSS1150-0023  & 11 50 43.88 & $-$00 23 54.07 & 17.00 & 1.9804 $\pm$ 0.0002 & 1.97987 $\pm$ 0.00002   &   $-$29.34 &   $-$29.09      \\
SDSS1219-0100  & 12 19 40.37& $-$01 00 07.49& 16.82 & 1.5770 $\pm$ 0.0002  & 1.5831 $\pm$ 0.0003  &   $-$29.03 &  $-$28.76           \\
SDSS1236-0331  & 12 36 02.34 & $-$03 31 29.94 & 16.91 & 1.8239 $\pm$ 0.0002   & 1.8464 $\pm$ 0.0007   &   $-$29.42 &   $-$29.14        \\
SDSS1354-0013  & 13 54 25.24 & $-$00 13 58.06 & 16.68 & 1.5124 $\pm$ 0.0002 & 1.5124 $\pm$  0.0001    &   $-$29.15 &  $-$28.90       \\
SDSS1431+0535  & 14 31 48.09 & $+$05 35 58.10 & 16.74 & 2.0964 $\pm$ 0.0002 & 2.09985 $\pm$  0.00007   &   $-$29.56 &  $-$29.32     \\
SDSS1437-0147  & 14 37 48.29 & $-$01 47 10.79 & 15.44 & 1.3091 $\pm$ 0.0001 & 1.30901 $\pm$  0.00004   &   $-$29.53 &   $-$29.38     \\

\hline
\hline
\end{tabular}
\end{center}
\noindent{

$^{(a)}$ Taken from SDSS .
$^{(b)}$ The fit method is described in the text. 
}

\end{table*}

%% file: tabs/comparison.tex
\begin{table}
\centering
\begin{center}
\caption{Power law slopes from different composites.}
\tablabel{comparison}
\begin{tabular}{cc}
\hline
\noalign{\smallskip}
Reference &  FUV slope, $\alpha$$^{(a)}$ \\  
\hline

This work  & $-1.70$   \\
Lusso et al. 2015  & $-1.39$   \\
Telfer et al. 2002  & $-1.31$   \\
Francis et al. 1991  & $-1.68 $   \\

Vanden Berk et al. 2001  & $-1.56$   \\
Glikman et al. 2006 & $-(0.45 - 1.63)$ $^{(b)}$  \\

\hline
\hline
\end{tabular}
\end{center}
\noindent{
$^{(a)}$ The slope of a power law in the Far UV.
$^{(b)}$ The range of slopes is due to different combination methods and regions selected for a power law fit. 
}

\end{table}

%% file: tabs/combinations.tex
\begin{table}
\centering
\begin{center}
\caption{Power law slopes from different combination methods.}
\tablabel{comb_meth} 
\begin{tabular}{cc}
\hline
\noalign{\smallskip}
Combination method &  $\alpha$$^{(a)}$ \\  
\hline

Weighted arithmetic mean  & $-1.70\pm 0.01$   \\
Arithmetic mean  & $-1.72\pm 0.01$   \\
Geometric mean  & $-1.71\pm 0.01$   \\
Median  & $-1.72\pm 0.01$   \\


Individual mean$^{(b)}$  & $-1.71\pm 0.09$   \\
Individual median & $-1.68$   \\

\hline
\hline
\end{tabular}
\end{center}
\noindent{
$^{(a)}$ The slope of a power law.
$^{(b)}$ The arithmetic mean of the slopes fitted in the individual spectra. The reported errors are the standard deviation of the individual slopes.

}

\end{table}

%% file: tabs/example.tex
\begin{table}
\centering
\begin{center}
\caption{Quasar composite spectrum.}
\tablabel{example}
\begin{tabular}{ccc}
\hline
\hline
\noalign{\smallskip}
$\lambda$ &  Rescaled flux density & Rescaled flux density error \\  
({\AA}) & F$_{\lambda}$ & $\delta$F$_{\lambda}$ \\
\hline

1000.1 & 14.2239 & 2.5207 \\
1000.5 & 14.4227 & 2.5335 \\
1000.9 & 14.5644 & 2.4424 \\
1001.3 & 14.6599 & 2.4170 \\
1001.7 & 14.7200 & 2.4097 \\
1002.1 & 14.7555 & 2.3526 \\
1002.5 & 14.7771 & 2.3333 \\
1002.9 & 14.7959 & 2.2527 \\

\hline
\hline
\end{tabular}
\end{center}
\noindent{
\tab{example} is available in its entirety at \url{https://github.com/jselsing/QuasarComposite} and in the electronic edition of the paper.

}

\end{table}